# Control of magnetization reversal in oriented Strontium Ferrite thin films.

Debangsu Roy* and P S Anil Kumar

Department of Physics, Indian Institute of Science, Bangalore 560012, India

Oriented Strontium Ferrite films with the c axis orientation were deposited with varying oxygen partial pressure on $Al_2O_3$(0001) substrate using PLD technique. The angle dependent magnetic hysteresis, remanent coercivity and temperature dependent coercivity had been employed to understand the magnetization reversal of these films. It was found that the Strontium Ferrite thin film grown at lower (higher) oxygen partial pressure shows Stoner-Wohlfarth type (Kondorsky like) reversal. The relative importance of pinning and nucleation processes during magnetization reversal is used to explain the type of the magnetization reversal with different oxygen partial pressure during growth.

*debangsu@physics.iisc.ernet.in

In today's world permanent magnets become ubiquitous in terms of its application in the modern technologies like motors, actuators, sensors, holding devices, ultrahigh density recording etc[1]. The thin magnetic films with strong perpendicular anisotropy are central to the investigation of the different applications. This is due to the reduction of the demagnetizing field present in the system, thus allowing sharp magnetic transition from one direction to the other. The response of a magnetic material in the external magnetic field can be understood if one considers the underlying magnetization reversal process in the system. In this regard the realization of the coercivity mechanism in the perpendicularly magnetized thin film system becomes important. The functionality of theses system can be tailored by considering the relative importance of the nucleation and pinning, which governs the coercivity mechanism in these systems. In the literature there are numerous reports about the investigation of the coercivity mechanism of the several hard magnetic material thin films most of which contains at least one of the rare earth element like Sm, Nd, Pr etc. [2-5] In this letter we have made an attempt to understand the magnetization reversal and the coercivity mechanism of the perpendicularly magnetized oriented hexagonal Strontium Ferrite thin film.

In the literature, there are numerous articles on the thin films of hexagonal ferrites like Barium Ferrite and Strontium Ferrites.[6, 7-9] Several techniques such as Pulsed Laser Deposition (PLD)[8, 9], sputtering [10], and plasma spraying [11] are used for the deposition of hexagonal ferrite thin films. Out of all these techniques, the PLD has emerged as an effective deposition technique for the oriented or epitaxial growth of the oxide hexagonal ferrite thin film[8, 9]. In the PLD technique a large number of processing parameters determine the crystalline quality of the deposited thin film which in turn governs the coercivity mechanism of the studied system. In this letter we report about the growth of the Strontium Ferrite thin film with a single orientation along c-axis on c-plane alumina by PLD and subsequently investigate the magnetization reversal mechanism with different oxygen partial pressure during the growth. However there is a lack of understanding about magnetization reversal mechanism in the oriented Strontium Ferrite thin film. It has been found that depending on the oxygen partial pressure, the deposited thin films of Strontium Ferrite exhibit different magnetization reversal mechanism and thus exhibit different coercivity value. Thus the coercivity mechanism in these Strontium

Ferrite thin films are investigated by the angular dependent and temperature dependent magnetic measurements and analyzed in accordance to the S-W model[12], Kondorsky model[13] and micromagnetic model[14]. The relative importance of the pinning and nucleation present in these systems are further discussed in the article along with the structural information analyzed using thin film XRD.

The Strontium Ferrite thin film of composition $SrFe_{12}O_{19}$ was deposited on the c-plane alumina substrate (0001) using Pulsed Laser Deposition technique in a HV chamber (base pressure < $2X10^{-5}$ mbar). During pulsed laser deposition, the substrate temperature was maintained at $800^{\circ}C$. In this article, we discuss the results of the Strontium Ferrite thin film which was deposited at two different oxygen partial pressure of 0.1 mbar and 0.4 mbar, retaining the substrate temperature at $800^{\circ}C$. The thin film of Strontium Ferrite deposited at 0.1 mbar and 0.4 mbar $O_2$ partial pressure has been named as SFO-LO and SFO-HO respectively. The thin film of Strontium Ferrite was further annealed in-situ at $500^{\circ}C$ at 500 mbar oxygen pressure for one hour to avoid the oxygen deficiency in the deposited film. It has to be noted that 4000 number of laser pulses with 300 mJ energy having fluence 2.5 $J/cm^2$ was used to have a thickness of ~ 20 nm for the Strontium Ferrite thin film. The crystallographic structure, growth and the quality of the thin Strontium Ferrite films deposited at two different oxygen partial pressure are analyzed by thin film x-ray diffraction and ω-scan about a particular Strontium Ferrite peak using Bruker D8 discover with Cu-Kα source. The magnetization measurement is performed at Lakeshore 7404 VSM with angular variation set up and PPMS set up by Quantum Design which operates in field upto 14T.

We have used X-ray diffraction studies to understand the structural orientation of the deposited Strontium Ferrite on the (0001) oriented alumina. The alumina is having trigonal crystal structure with the lattice parameter of a= 4.754 Å and c= 12.99 Å. The lattice parameters for the Strontium Ferrite are a= 5.882 Å and c= 23.023 Å. Figure 1 shows the typical θ-2θ XRD scan of SFO-HO and SFO-LO thin film ($SrFe_{12}O_{19}$ (20 nm)/ $Al_2O_3$). It is evident from the figure 1 that the X-ray diffraction study of SFO-HO and SFO-LO thin film indicates the presence of the strong (000*l)* reflection of Strontium Ferrite indicating c-axis oriented growth of the Strontium Ferrite on c-plane alumina substrate. This is in accordance to the fact that the hcp oxygen plane of the Strontium Ferrite aligns preferentially along the hcp oxygen plane of the alumina.

The peak which was marked as (**) corresponds to the peak intrinsic to the instrument. Thus the X-ray diffraction pattern for the Strontium Ferrite thin film on c-plane alumina substrate confirms the oriented growth of the Strontium Ferrite along (000*l*) direction. In order to understand the relative importance of the mosaicity and the crystalline quality of SFO-HO and SFO-LO, we have undertaken XRD ω-scan about the Strontium Ferrite (00014) reflection as shown in the inset of the figure 1. Generally the mosaicity of a crystal is the width of the distribution of the mis-orientation angle of the crystal plane in the crystal and is directly related to the crystalline quality. The presence of the mosaic defects in the crystal corresponds to the broadening of the rocking curve. Thus the estimation of the width i.e. the FWHM of the rocking curve for both SFO-HO and SFO-LO is a measure of the crystalline quality of the thin film. The inset of the figure 1 shows the Gaussian fitting of the rocking curve for both SFO-HO and SFO-LO and it was found that the peak position remains same in both the cases. The FWHM for SFO-HO and SFO-LO can be estimated as $0.231^0$ and $0.14^0$. The FWHM for the c-plane alumina obtained from the rocking curve about (0006) reflection of the substrate can be calculated as $0.0143\ ^0$. Thus the lesser value of FWHM for the SFO-LO compared to SFO-HO corresponds to the better crystallinity in SFO-LO compared to SFO-HO.

The figure 2 shows the magnetic hysteresis loop for SFO-LO corresponding to the magnetic field applied at different angle Φ, with respect to the film normal direction ((0001) direction of the c-plane $Al_2O_3$ substrate). All the magnetic measurements were done in a Lakeshore 7404 VSM. The image inside the figure 2 shows the angle Φ between the applied field and the film normal. In addition, the angular dependence of the magnetization of SFO-HO exhibits correspondence to the magnetization behaviour of SFO-LO with respect to the (0001) direction of the c-plane alumina substrate at different angles. (Supplement 1).

It has been found that for both the film, the coercivity value decreases as the angle between the applied magnetic field and the film normal increases from $0^0$ to $90^0$. From the magnetization measurement for both the thin film, we can conclude that both the film exhibit perpendicular anisotropy corresponding to the $\Phi = 0^0$ configuration. The magnetization behaviour at $\Phi = 90^0$ for both the thin film shows a characteristics of the magnetic hysteresis loop along the hard axis. Thus the deposited Strontium Ferrite on c-plane alumina, is having the easy axis of magnetization along the c axis of the Strontium Ferrite and the corresponding hard axis lies in the film plane. The inset of the figure 2 illustrates the variation of the coercivity of SFO-HO and SFO-LO with different angle between applied field and (0001) direction of the substrate. It has also been found from the inset of the figure 2 that the coercivity for SFO-HO and SFO-LO are ~2600 Oe and ~3820 Oe respectively along the easy direction of the magnetization. The coercivity of the SFO-LO is always higher compared to SFO-HO for all the angles Φ. Generally in the thin film of hard magnetic material, domain forms with the application of the reverse magnetic field[1]. Subsequently the movement of the domain walls determine the coercivity of the system. However the motion of this domain wall is affected by the anisotropy of the system, density of pinning state, surface asperities etc. Thus for better understanding of the coercivity mechanism in SFO-HO and SFO-LO, all the above mentioned parameter should be taken into account.

In order to understand the role of the defect site in the coercivity mechanism for SFO-HO and SFO-LO, we have obtained DC demagnetization remanence for the film at various angles between the applied field and the easy direction of the magnetization using Lakeshore 7404 VSM. From the corresponding DC demagnetization remanence curve we have calculated the corresponding remanent coercivity $H_{cr}$ .The corresponding field value for which the remanent magnetization becomes zero can be defined as the remanent coercive field. This depends only on the irreversible part of the magnetization and independent of the magneto-static interaction in the system as the net magnetization is zero at this point on the reversal curve. Using this procedure, we have calculated the corresponding remanent coercive field value for both SFO-HO and SFO-LO at various angles Φ. Generally the simplest way to explain the angular dependence of the remanent coercivity is attributed to Kondorsky model[13] and the same can be expressed as $H_{cr}(\Phi)/H_{cr}(0) = 1/\cos\Phi$. Here Φ is the angle between the film normal and the applied field as denoted by the diagram in the figure 2. Here $H_{cr}(0)$ corresponds to the remanent coercivity along the easy direction of the magnetization and the expression for the Kondorsky model has been normalized using the value of $H_{cr}(0)$. According to this model, the magnetization reversal occurs through the domain wall formation and its subsequent motion. The physical picture for this model can be visualized to the situation when an external magnetic field is applied to the thin film at an arbitrary angle apart from the easy direction of the magnetization. In this scenario, if the applied magnetic field is insufficient to rotate the magnetization of the film from the easy axis direction, then the component of the applied magnetic field along the easy direction of the magnetization can push the domain wall from one meta-stable energy minimum to the other. According to the Kondorsky model the obtained $H_{cr}$ should always be lesser than the corresponding anisotropy field of the system. The Kondorsky model is well applicable for the switching of the longitudinal recording media.[1]

Another model which explains the variation of the remanent coercivity is the Stoner Wohlfarth (S-W) Model[12]. This model considers the coherent rotation as the process for magnetization reversal for the uni-axial, single domain non-interacting particles. The S-W switching field can be expressed as $H_{sw}(\Phi) / H_{sw}(0) = 1 / \left(\cos^{2/3}\Phi + \sin^{2/3}\Phi\right)^{3/2}$, here $H_{sw}(0)$ is equal to the anisotropy field of the system. In the present context, we have considered that the remanent switching field is as same as remanent coercivity of the system. Thus in this model the remanent coercivity becomes maximum when the applied magnetic field lies either parallel to the easy direction of the magnetization or to the hard direction. The remanent coercivity according this model is found to be minimum at the angle $\Phi = 45^0$.

Figures 3(a) and 3(b) illustrate the variation of the remanent coercivity for SFO-LO and SFO-HO. The variation of the simulated $H_{cr}/H_{cr}(0)$ of the S-W model and the Kondorsky model with the different angle $\Phi$ has also been shown in the figure 3(a) and 3(b) respectively. It has been found from the figure 3(a) that for SFO-LO, the angular dependence of the reversal resembles the functional form of the S-W model. It has also been found that the minimum of the normalized remanent coercivity occurs at $45^0$ for SFO-LO which is identical to the angle at which switching field becomes minimum for the S-W model. This suggests that the magnetization reversal is coherent in nature for the hard SFO-LO. However, we have found that the magnitudes of the remanent coercivity for SFO-LO differs from the value predicted from the S-W model for different angle $\Phi$ w.r.t. film normal. But the nature of the variation is same in both the cases. In addition, the depth of the minimum of the normalized remanent coercive field is less compared to the depth of the normalized remanent coercive field at an angle $45^0$ as predicted by the S-W model. This deviation from the ideal S-W model for the 20 nm thick Strontium Ferrite film can be explained by considering the existence of the large grain of the Strontium Ferrite. The large grain will introduce the domain wall inside the grain and thus the coherence of the magnetization in SFO-LO gets reduced. However, we also have to consider the presence of the defects for the understanding of the magnetization reversal in SFO-LO. Thus the apparent contradiction of the magnetization reversal in the hard magnetic SFO-LO which shows S-W like magnetization reversal behaviour can be explained according to a two step process of magnetization switching.[15] The sanity of this model is further verified from the other studies that will be discussed in the later part of this article. According to this process the reverse domain forms by nucleation and subsequently these domains propagate. So the relative importance of the nucleation field and the domain wall propagation field controls the switching behaviour in the system. Generally the existence of the defects in the system controls the domain wall propagation field as these defects acts as a pinning centre. So when the nucleation field value is higher compared to the domain wall propagation field value, magnetization does not flip until the applied field overcomes the nucleation field value. As soon as the domain gets nucleated, the domain wall forms and propagates resulting in the variation of the remanent coercive field with the angle $\Phi$ resembling S-W like switching. Thus in SFO-LO, the magnetization switching mechanism is nucleation dominated and the field at which the nucleation occurs is higher than the corresponding domain wall propagation field.

It has been found that the angular variation of the normalized remanent coercivity for SFO-HO exhibit Kondorsky type reversal. But after an angle $\Phi > 30^0$, the angular dependence of the normalized remanent coercive field value deviates from the remanent coercive field value obtained from the simulated Kondorsky model. This deviation can also be explained by considering the two step magnetization reversal process[15] which has been used for the understanding of the magnetization switching in SFO-LO. In this system if the nucleation field is lesser than the domain wall propagation field, the magnetization of the system undergoes nucleation at the beginning. But the magnetization does not flip with the application of the reverse field, until the energy corresponding to the domain wall propagation field is overcome. In this case the magnetization reversal is controlled by the Kondorsky model. The higher concentration of the defects present in SFO-HO compared to the SFO-LO which is evident from the inset of the figure 1 corresponds to the higher domain wall propagation field compared to the nucleation field in SFO-HO. Thus in this system the magnetization reversal is primarily controlled by the domain wall motion but the rotation mode owing to the nucleation also exists. This two step model also corroborates well to the fact that the coercivity is higher in SFO-LO compared to SFO-HO as the nucleation field value is higher in SFO-LO compared to the domain wall propagation field in SFO-HO. This result in a lesser coercivity in SFO-HO compared to SFO-LO. In the present context of the coercivity mechanism, we have considered negligible effect of the thermally activated magnetization reversal process which is relevant in some particular system.

In order to understand the quantitative importance of the magneto-static interaction owing to the size of the defects and the respective pinning or nucleation strength which together determines the coercivity of the system, a micromagnetic model has been used[14]. According to this model, the coercivity of the system can be expressed as $H_c = \alpha_{eff} H_n(T) - N_{eff} M_s(T)$.[16] Here the coercivity $H_c$ is described as the reduction of the nucleation field $H_n$ owing to the magneto-static interaction $N_{eff} M_s$. Theoretically the nucleation field can be expressed as the field for which the magnetization reversal would take place for the SW particle and it is an intrinsic property of the material. But in reality, the magnetization reversal is affected by the misalignment of the grains, surface defects, stray field (magneto-static interaction), structural inhomogenities etc. Generally the defects are the source for the stray field as the magnetic induction creates a strong local demagnetizing field in the vicinity of these defects. These defect can acts as a nucleation or pinning centre depending on their respective size and thus become a major parameter for deciding the coercivity of the system. These extra effects have been included in the expression of coercivity as an extra parameter $\alpha_{eff}$ and $N_{eff}$ which are temperature independent. The parameter $\alpha_{eff}$ stands for the microstructure related dependence factor and can further be split as $\alpha_{eff} = \alpha_k \alpha_\Phi$ where the corresponding parameter $\alpha_k$ is Kronmuller parameter [16-18] which is related to the structure of the sample. The parameter $\alpha_k$ is also independent of the temperature in the present context as the size of the microstructure of the deposited hard Strontium Ferrite does not change with the temperature. The other parameter $\alpha_\Phi$ depends on the easy axis misorientation. The demagnetizing parameter $N_{eff}$ depends on the distribution of the grain shape in the material, structural defect owing to the growth of the thin film etc.

To estimate the parameter $N_{eff}$ and the $\alpha_{eff}$ , we have done hysteresis measurement for SFO-HO and SFO-LO at different temperatures starting from 300K to 5K with an interval of 25K at Quantum Design PPMS. During this measurement, the magnetic field was always applied along the c-axis of the Strontium Ferrite i.e. along the easy axis of magnetization. It has been found that, the shape of the hysteresis does not change as the temperature has been varied from 300K to down 5K, apart from the individual change in saturation magnetization and coercivity for SFO-HO and SFO-LO. The maximum applied field is 14T and the saturation magnetization has been measured at 14T after eliminating the diamagnetic contribution of the substrate. It has to be noted that during this measurement configuration when the magnetic field is applied along the easy direction of the magnetization of SFO-HO and SFO-LO, the magnitude of the parameter $\alpha_{\Phi}$ can be considered as unity.

According to literature [19], the nucleation field in bulk Strontium Ferrite crystal can be determined by the first anisotropy constant $K_1$ and the nucleation field can be expressed as $H_n = 2K_1/M_s$ . So after incorporating the value of the nucleation field in the micromagnetic equation, the resultant equation can be written as

$$H_c(T)/M_s(T) = \alpha_K \, 2K_1/M_S^2(T) - N_{eff} M_s \quad \textbf{(1)}$$

Figure 4(a) and 4(b) shows the variation of the $H_c(T)/M_s(T)$ vs. $2K_1/M_S^2(T)$ for SFO-LO and SFO-HO. It has to be noted that, we have considered the first anisotropy constant value of $3.5\times10^6$ erg/cc for the calculation of the nucleation field. We have done the linear fitting of the variation of the $H_c(T)/M_s(T)$ vs. $2K_1/M_S^2(T)$ using the equation (1). The obtained fitting parameter are $\alpha_k$= 0.103±0.009, $N_{eff}$ = 1.753±0.258 and $\alpha_k$= 0.147±0.005, $N_{eff}$ = 2.904±0.219 for SFO-HO and SFO-LO respectively. Generally the demagnetization parameter $N_{eff}$ can attain values between 0 and 1 for the homogeneously magnetized sample. But for both the sample SFO-HO and SFO-LO, we have obtained the value of the $N_{eff}$ higher than 1. This corresponds to the presence of the enhanced stray field, which is higher than the saturation magnetization of the magnetic phase. The corresponding higher value of the $N_{eff}$ i.e. the measure of the stray field for SFO-LO compared to SFO-HO indicates that the defect size is higher in SFO-LO compared to SFO-HO. During the deposition of the Strontium Ferrite thin film on c-plane alumina substrate, the growth mode becomes important from structural as well as magnetic point of view. Generally during pulsed laser deposition of the oxide thin film, the layer by layer growth is desirable as only then one can obtain atomically smooth surface and superior physical property. Generally the high temperature of the substrate along with the low oxygen pressure during deposition, enhance the mobility of the adatoms on the substrate thus increasing the possibility of the 2D layer by layer growth. As a result the probability of nucleation of the next layer on the top of 2D island becomes minimum and the growth of the next layer is only possible after the completion of the previous layer. However, the higher oxygen pressure during deposition leads to a multilevel growth. In the present context of SFO-HO and SFO-LO, one would expect multilevel growth mode for SFO-HO leading to higher structural defects and pinning sites. But for SFO-LO the concentration of the defect is less as evident from the rocking curve measurement in the inset of the figure 1. It can also be concluded that the size of the defect is much larger than the domain wall width of the system; otherwise SFO-LO would exhibit a pinning dominated magnetization reversal process. Thus these defects in SFO-LO acts as a nucleation centre whereas the defects present in SFO-HO pinned the magnetization in the system. According to the literature the critical value of 0.35 [18] of the microstructure parameter $\alpha_k$ corresponds to the nucleation dominated magnetization reversal. The present value of the $\alpha_k$ confirms that both the pinning and nucleation coexist for both SFO-HO and SFO-LO. However, the lesser value of $\alpha_k$ for SFO-HO than SFO-LO indicates that the pinning is stronger in SFO-HO in comparison with SFO-LO. This is in well agreement to the previous findings regarding the quantitative estimation of the magneto-static interaction.

We have presented a detailed structural analysis using thin film XRD, angle dependent magnetic hysteresis and remanent coercivity measurement and coercivity mechanism by micromagnetic analysis for SFO-HO and SFO-LO. We have initially deposited SFO-LO and SFO-HO on (0001) alumina, with different oxygen pressure of 0.1 mbar and 0.4 mbar keeping the substrate temperature at 800$^{\circ}$C. The rocking curve analysis of the oriented SFO-HO and SFO-LO reveals the better crystallinity in SFO-LO compared to SFO-HO.

The angular dependence of the coercivity for both the SFO-HO and SFO-LO indicates that the deposited films are having easy axis of magnetization along the growth direction i.e. along the film normal direction. Correspondingly it was found that the hard axis lies in the film plane. We have found that there is a significant variation of the coercivity between SFO-HO and SFO-LO. We have made an attempt to understand the factors governing the coercivity mechanism in these films by considering the pinning and nucleation centres. It has been found that for SFO-HO, the coercivity mechanism follows a pinning dominated process where the field corresponding to the domain wall propagation is higher than that of the nucleation. But for SFO-LO, the coercivity mechanism is governed by nucleation of the domains and corresponding motion of the domain walls. The micromagnetic analysis of the coercivity in SFO-HO and SFO-LO confirms that the size of the defects is bigger in SFO-LO compared to SFO-HO. However, it has also been confirmed that the density of the defects are more in SFO-HO compared to SFO-LO. The quantitative information regarding the strength of the magneto-static interaction for SFO-HO and SFO-LO reveals that, the strength of the nucleation field in SFO-LO is more compared to the strength of the domain wall propagation field which governs the coercivity in the respective systems.

List of Figures:

**Figure 1:** θ-2θ XRD pattern for $SrFe_{12}O_{19}$ (20 nm) film grown on $Al_2O_3$ (0001) with different oxygen concentration. The red and black coloured graph denotes the $SrFe_{12}O_{19}$ thin film deposited at 0.4 and 0.1 mbar $O_2$ pressure. The (**) peak corresponds to the intrinsic peak of the instrument. (inset) Rocking curve and the corresponding Gaussian fitting for sample SFO-HO and SFO-LO.

**Figure 2:** Magnetization vs. Applied Field for SFO-LO at various applied field angle with respect to the (0001) direction of the c-plane alumina substrate. (inset):Variation of the coercivity of SFO-HO and SFO-LO with angle between applied magnetic field and film normal.

**Figure 3(a):** Variation of the normalized remanent coercivity for SFO-LO with the angle Φ. **(b):** Variation of the normalized remanent coercivity for SFO-HO with the angle Φ.

**Figure 4 (a):** Variation of the $H_c(T)/M_s(T)$ vs. $2K_1/M_S^2(T)$ for SFO-LO. The dotted line shows the corresponding linear fit according to the eq$^n$(1). **(b):** Variation of the $H_c(T)/M_s(T)$ vs. $2K_1/M_S^2(T)$ for SFO-HO. The dotted line shows the corresponding linear fit according to the eq$^n$(1).

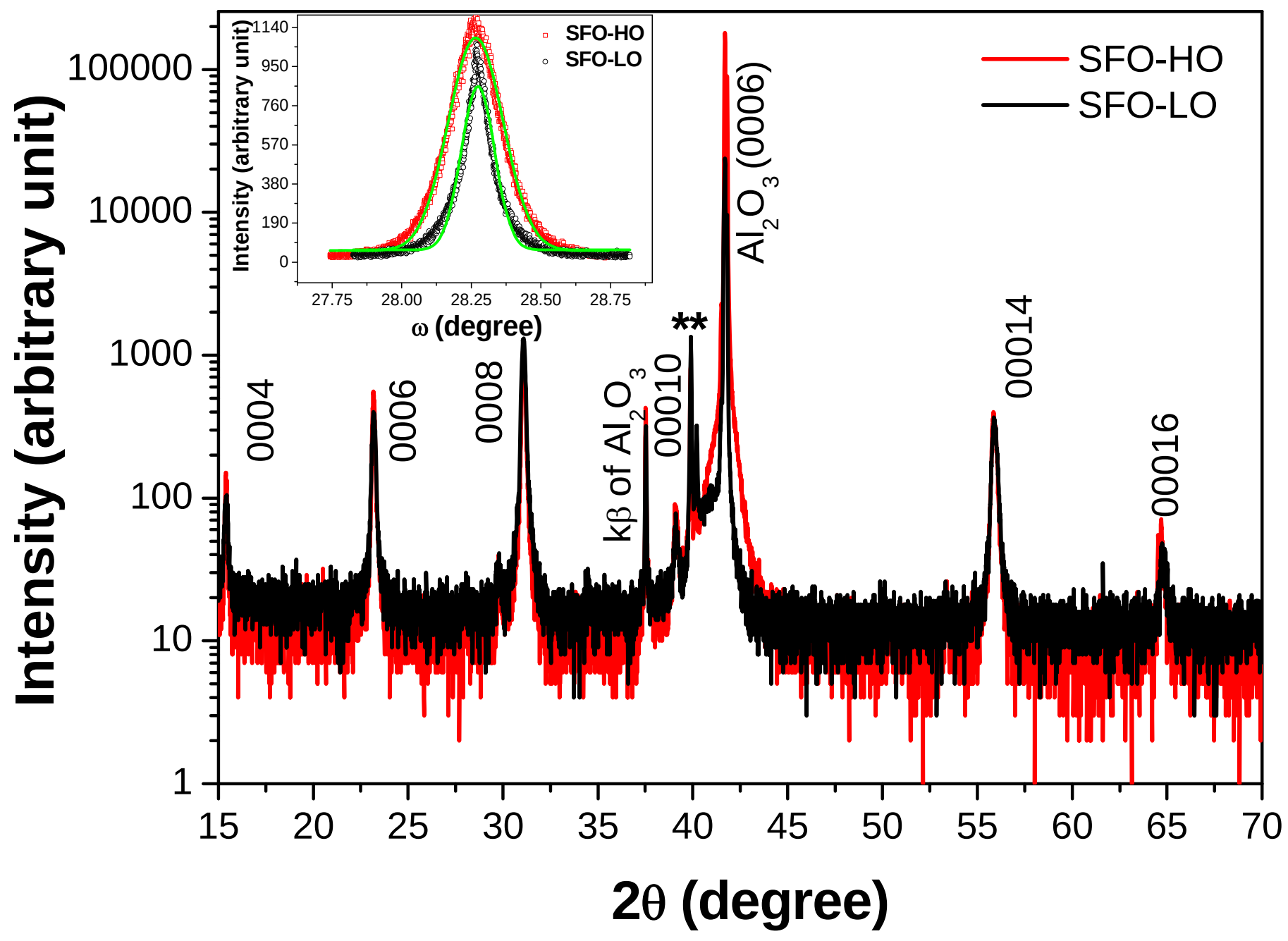

Figure 1

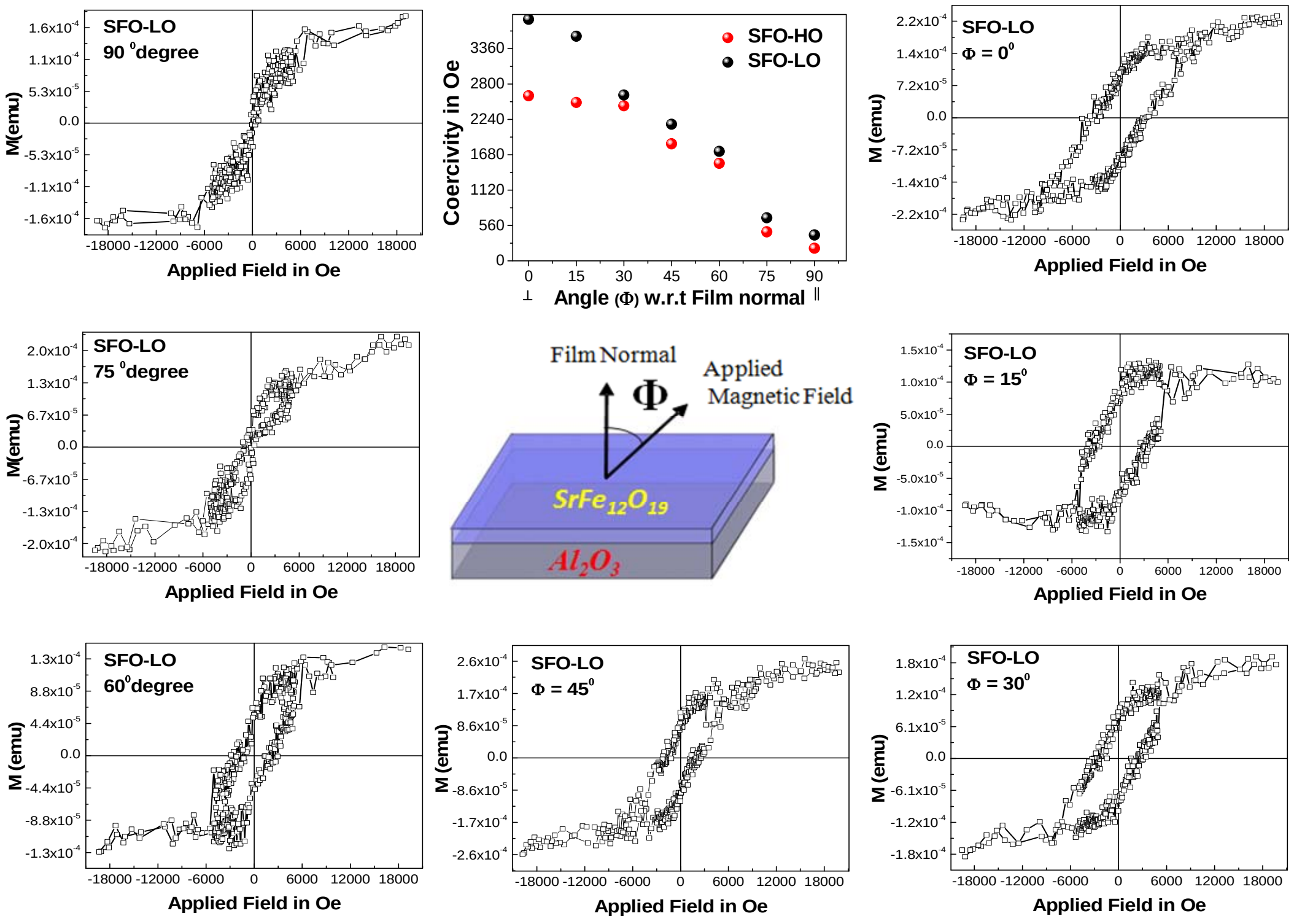
SFO-LO
90 °degree
M(emu)
Applied Field in Oe
Coercivity in Oe
SFO-HO
SFO-LO
Angle (Φ) w.r.t Film normal
SFO-LO
Φ = 0°
M (emu)
Applied Field in Oe
SFO-LO
75 °degree
M(emu)
Applied Field in Oe
Film Normal
Φ
Applied
Magnetic Field
SrFe12O19
Al2O3
SFO-LO
Φ = 15°
M (emu)
Applied Field in Oe
SFO-LO
60°degree
M (emu)
Applied Field in Oe
SFO-LO
Φ = 45°
M (emu)
Applied Field in Oe
SFO-LO
Φ = 30°
M (emu)
Applied Field in Oe

Figure 2

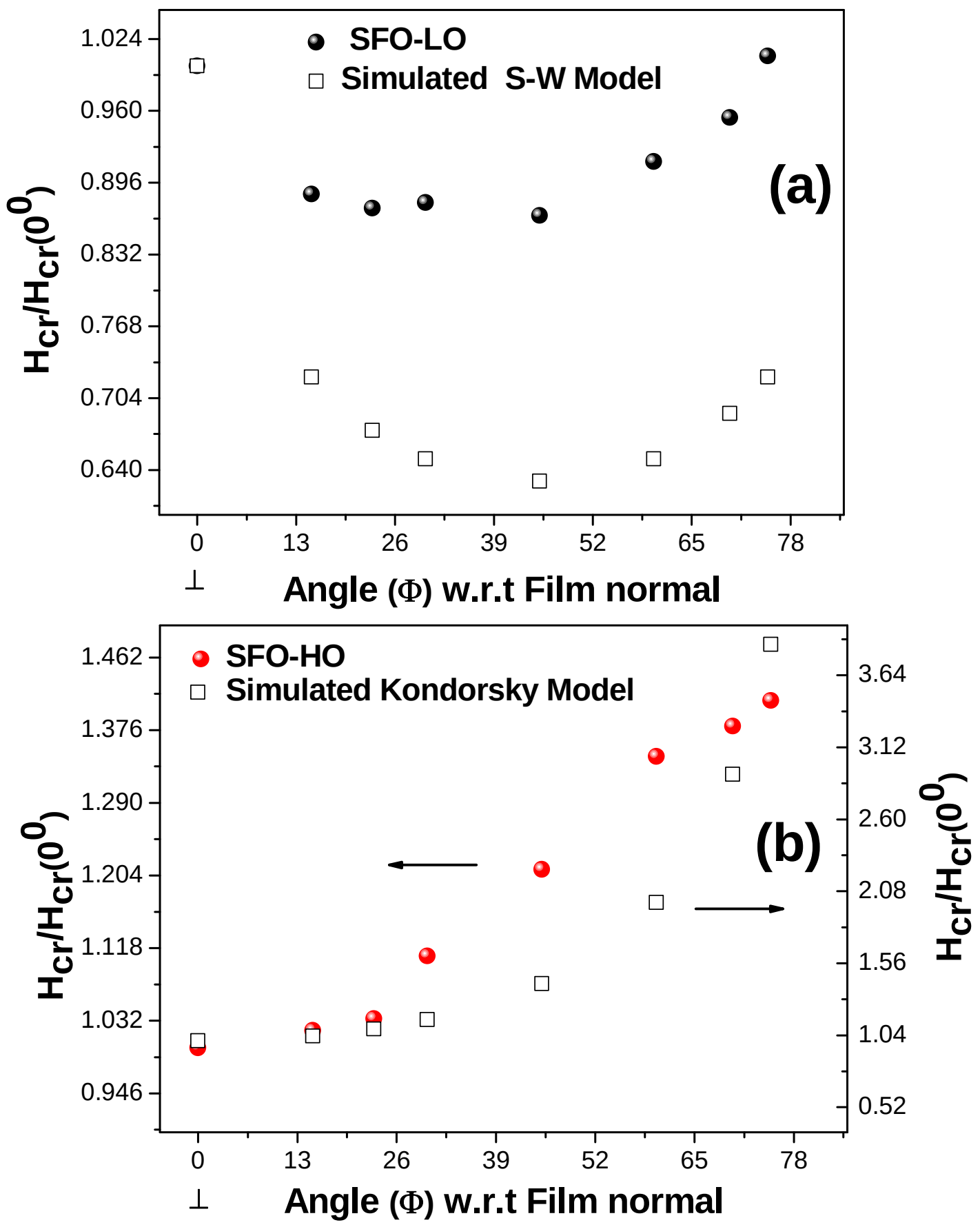

Figure 3

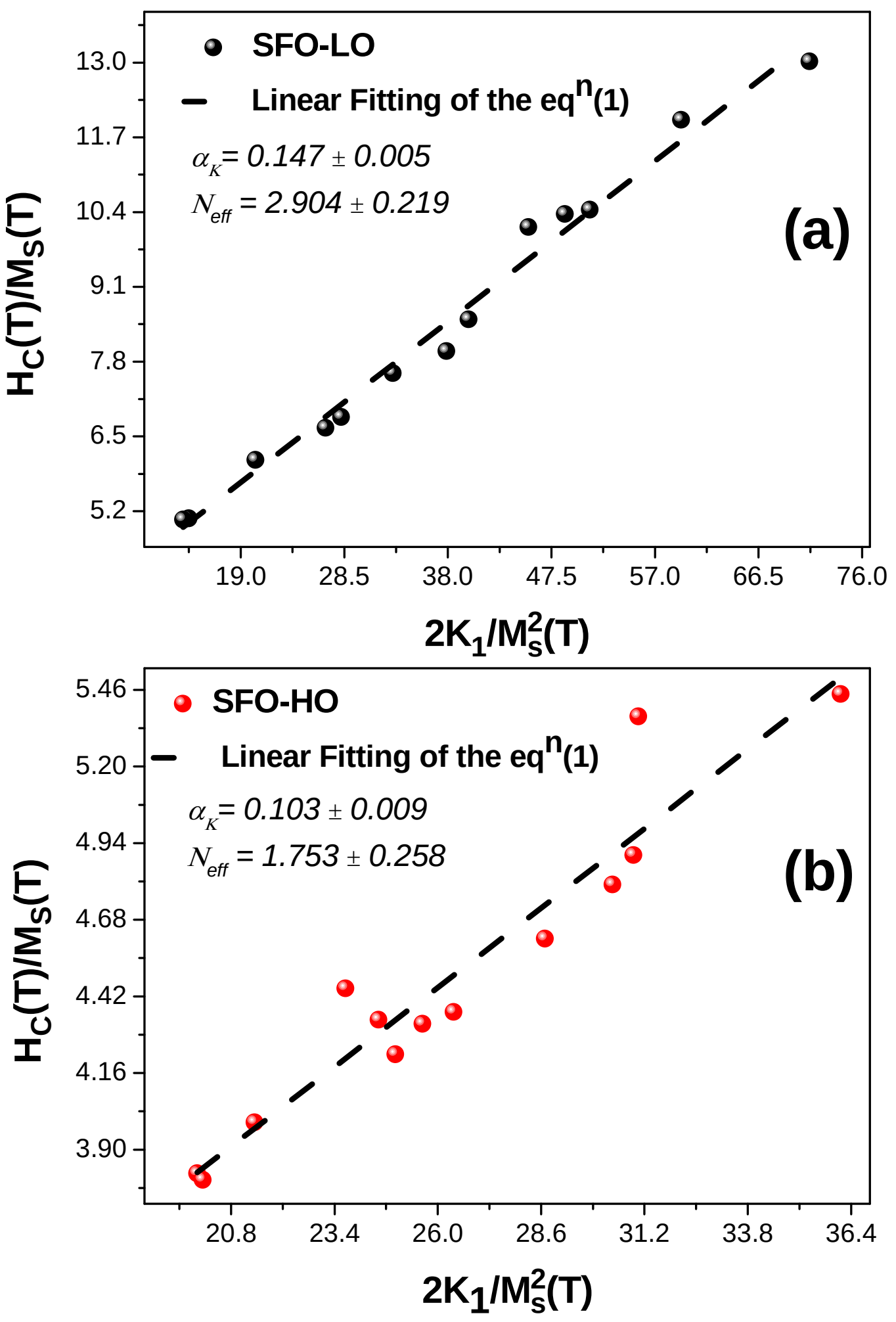

Figure 4